\documentclass[showpacs]{revtex4}
\usepackage{graphicx,amsmath,amssymb,amsfonts,bbm,latexsym,color,dcolumn,epsf}

\def\be{\begin{equation}}
\def\ee{\end{equation}}
\def\lp{\left(}
\def\rp{\right)}
\def\lb{\left[}
\def\rb{\right]}
\def\om{\omega}

\def\ck{\chi_k}

\begin{document}

\title{New counterterms induced by trans-Planckian physics in semiclassical gravity}
\author{D. L\'opez Nacir \footnote{dnacir@df.uba.ar}}
\author{F. D. Mazzitelli \footnote{fmazzi@df.uba.ar}}
\affiliation{Departamento de F\'\i sica {\it Juan Jos\'e
Giambiagi}, Facultad de Ciencias Exactas y Naturales, UBA, Ciudad
Universitaria, Pabell\' on I, 1428 Buenos Aires, Argentina}

\begin{abstract}
We consider free and self-interacting quantum scalar fields
satisfying modified dispersion relations in the  framework of
Einstein-Aether theory. Using adiabatic regularization,  we study
the renormalization of the equation for the mean value of the
field in the self-interacting case, and the renormalization of the
semiclassical Einstein-Aether equations for free fields. In both
cases we consider Bianchi type I background spacetimes. Contrary
to what happens for {\it free} fields in {\it flat}
Robertson-Walker spacetimes, the self-interaction and/or the
anisotropy produce non-purely geometric terms in the adiabatic
expansion, i.e terms that involve both the metric  $g_{\mu\nu}$
and the aether  field $u_{\mu}$. We argue that, in a general
spacetime, the renormalization of the theory would involve new
counterterms constructed with $g_{\mu\nu}$ and $u_{\mu}$,
generating a fine-tuning problem for the Einstein-Aether theory.
\end{abstract}


\maketitle

\section{Introduction}

In the last years it has been realized that the (still unknown)
physics at very high energies may not be inaccessible from an
observational point of view. Indeed, trans-Planckian physics may
have left  an imprint in the inhomogeneities of the cosmic
microwave background radiation \cite{CMBtrans}, in the evolution of the scale
factor of the universe \cite{scale}, in the propagation of gamma ray bursts \cite{gamma},
etc.

In the absence of a full theory, the theoretical approach to this
problem is phenomenological. One possibility, that we will
consider here, is to assume that the physics at high energies is
such that its main effect is a modification of the dispersion
relation of the quantum fields, thus violating Lorentz symmetry.
Although this is a  simplistic approach, it could be useful to
investigate whether the trans-Planckian effects could lead to
observable consequences or not in a given particular situation, by
testing the robustness of the results under changes in the
dispersion relations at very high energies.

The Modified Dispersion Relations (MDR) will obviously affect the
structure of the quantum field theory, in particular its
renormalizability. Having in mind applications to cosmology, in
previous papers \cite{NosUno,NosProc,NosDos}, we have analyzed in
detail the renormalization of free field theories with MDR in flat
Robertson Walker spacetimes. We have shown that the theory can be
renormalized using a generalization of the well known adiabatic
regularization \cite{ad-old,equiv} that is used in theories with
standard dispersion relations. As for the usual case, the
adiabatic expansion of the energy momentum tensor contains
divergent terms that can be written in terms of geometric tensors
in $n$-dimensions, and therefore the theory can be renormalized by
absorbing the infinities into the bare gravitational constants of
the theory. It is remarkable that this can be done whatever the
dispersion relation. This somewhat surprising result could be a
peculiarity of flat Robertson Walker metrics \cite{ted} and/or
valid only for free fields, and therefore it is of interest to
investigate more general situations.

In this paper we extend the adiabatic regularization to the case
of self-interacting fields and anisotropic metrics (Bianchi type
I). We will work within the context of the so called
Einstein-Aether theory \cite{Jacobson}, a covariant theory of gravity in
which the metric is coupled to a dynamical vector field. This
field,  that breaks Lorentz invariance dynamically, is also
coupled to the derivatives of the quantum matter fields, leading
to  MDR that contain higher powers of the momenta. The specific
model is introduced in Section II.

In Section III we consider a self-interacting scalar field on
Bianchi type I metrics and discuss the renormalization of the
equation for the mean value of the field $\phi_0$. In order to do
this, it will be necessary to compute the mean value of the
fluctuations of the field
$\langle\hat\phi^2\rangle=\langle(\phi-\phi_0)^2\rangle$. We will
calculate explicitly this quantity up to the second adiabatic
order and show that, contrary to what happens for the usual
dispersion relation, the second adiabatic order cannot be entirely
written in terms of the metric and its derivatives, but also
involve the aether field $u_{\mu}$  and its derivatives. This
property of  $\langle\hat\phi^2\rangle$ is valid even for free
fields in flat Robertson-Walker spacetimes.

In Section IV we analyze the renormalizability of the
Semiclassical Einstein-Aether Equations (SEAE) for the case of
free scalar fields with MDR in Bianchi type I universes. We
compute $\langle T_{\mu\nu}\rangle$ up to the second adiabatic
order. The zeroth adiabatic order is divergent whatever the
dispersion relation. Being proportional to $g_{\mu\nu}$, the
divergence can be absorbed into a redefinition of the comological
constant. The second adiabatic order is shown to be divergent for
dispersion relations that involve powers of the momenta  smaller
than or equal to four. This adiabatic order contains  a term
proportional to $G_{\mu\nu}$, that renormalizes Newton's constant.
However, it also contains an additional non-purely geometric term,
proportional to the variation of $(\nabla_{\mu}u^{\mu})^2$.  When
this term is divergent, a new counterterm has to be introduced to
renormalize the theory, even if originally not present in the
classical Lagrangian. On the other hand, if it is finite, a
counterterm would be necessary   to make the theory consistent
with observations.

In Section V we argue that, for a general metric, the
renormalization of the infinities produced by a quantum free field
satisfying MDR will induce all possible counterterms involving up
to two derivatives of the metric $g_{\mu\nu}$ and the vector
$u_{\mu}$. As shown in Ref. \cite{Jacobsondebil}, the coefficients
of terms like $(\nabla_{\mu}u^{\mu})^2$,
$R_{\mu\nu}u^{\mu}u^{\nu}$, etc,  are strongly constrained
observationally by post-Newtonian parameters, and therefore the
counterterms induced by trans-Planckian physics should be fine
tuned to satisfy these constraints.

Throughout the paper we set $c=1$ and adopt the sign convention
denoted (+++) by Misner, Thorne, and Wheeler \cite{MTW}.

\section{The Model}

We work in the frame of a generally covariant theory of gravity
coupled to a dynamical vector field $u^{\mu}$ that breaks local
Lorentz symmetry. The most general action that is quadratic in
derivatives is given by \cite{Jacobson}: \be S_{G}=\frac{1}{16\pi
G }\int d^n x \sqrt{-g} (R-2\Lambda+\mathcal{L}_{u}),\label{Sg}\ee
where $g=det(g_{\mu\nu})$, $R$ is the Ricci scalar, $\Lambda$ and
$G$ are the bare cosmological and Newton's constants, and
$\mathcal{L}_{u}$ describe the dynamics of the additional degree
of freedom $u^{\mu}$, \be\label{lu}
\mathcal{L}_{u}=-\tilde{\lambda}(g^{\mu\nu}u_{\mu}u_{\nu}+1)-b_1
F_{\mu\nu}F^{\mu\nu}-b_2 (\nabla_{\mu}u^{\mu})^2-b_3
R_{\mu\nu}u^{\mu}u^{\nu}-b_4
u^{\rho}u^{\sigma}\nabla_{\rho}u_{\mu}\nabla_{\sigma}u^{\mu},\ee
where $F_{\mu\nu}=\nabla_{\mu}u_{\nu}-\nabla_{\nu}u_{\mu}$. The
Lagrange multiplier $\tilde{\lambda}$ is introduced to impose the
condition $u_{\mu}u^{\mu}=-1$ and the coefficients $b_i$
($i=1,2,3,4$) are arbitrary. The term
$\nabla_{\mu}u_{\nu}\nabla^{\nu}u^{\mu}$ coincides with
$(\nabla_{\mu}u^{\mu})^2-R_{\mu\nu}u^{\mu}u^{\nu}$ up to a total
derivative, and hence has been omitted.

We consider a quantum scalar field $\phi$ with a generalized
dispersion relation propagating in a curved space-time with a
classical background metric given by \be
ds^2=g_{\mu\nu}dx^{\mu}dx^{\nu}\equiv
-(u_{\mu}dx^{\mu})^2+\perp_{\mu\nu}dx^{\mu}dx^{\nu},\ee where
$\mu,\nu= 0,1...n-1$ (with $n$  the space-time dimension) and
$\perp_{\mu\nu}\equiv g_{\mu\nu}+ u_{\mu} u_{\nu}$. The action for
the scalar field can be written as: \be S_{\phi}=\int d^n x
\sqrt{-g}
(\mathcal{L}_{\phi}+\mathcal{L}_{cor}+\mathcal{L}_{int}),\ee where
 $\mathcal{L}_{\phi}$ is the standard
Lagrangian of a free, massive, minimally coupled scalar field \be
\mathcal{L}_{\phi}=-\frac{1}{2}\lb g^{\mu
\nu}\partial_{\mu}\phi\partial_{\nu}\phi+m^2\phi^2\rb,\ee
 $\mathcal{L}_{cor}$ is the corrective lagrangian that gives rise
to a generalized dispersion relation \be
\mathcal{L}_{cor}=-\sum_{s,p} b_{sp}
(\mathcal{D}^{2s}\phi)(\mathcal{D}^{2p}\phi),\ee where  $0< p\leq
s$, $b_{sp}$ are arbitrary coefficients, and
$\mathcal{D}^{2}\phi\equiv\perp_{\mu}^{\lambda}\nabla_{\lambda}\perp_{\gamma}^{\mu}\nabla^{\gamma}\phi$
($\nabla_{\mu}$ is the covariant derivative corresponding to the
metric $g_{\mu\nu}$ and $\perp_{\mu}^{\lambda}\equiv
g^{\lambda\nu}\perp_{\mu\nu}$). The interaction Lagrangian
$\mathcal{L}_{int}$ contains the following terms:
\be\label{lintgen} \mathcal{L}_{int}=-\frac{1}{2}[\xi R+\xi_1
F_{\mu\nu}F^{\mu\nu}+\xi_2(\nabla_{\mu}u^{\mu})^2+\xi_3
\nabla_{\mu}u_{\nu}\nabla^{\nu}u^{\mu}+\xi_4
u^{\rho}u^{\sigma}\nabla_{\rho}u_{\mu}\nabla_{\sigma}u^{\mu}+\xi_5
u^{\mu}u^{\nu}R_{\mu\nu}]\phi^2-\lambda\phi^4,\ee where $\xi$,
$\xi_i$ ($i=1,2,3,4,5$) and $\lambda$ are bare parameters. Note
that, in addition to the self-interaction and the standard
coupling to the Ricci scalar, we have also included couplings
between $\phi^2$ and non-purely geometric terms that involve the
aether field $u_{\mu}$. Note also that, if we assume that the MDR
depart from the usual one at a given scale $M_C$, the coefficients
$b_{sp}$ scale as $b_{sp}\sim M_C^{2(1-s-p)}$.

In the rest of the paper we will consider a four-dimensional
Bianchi type I space-time with line element \be
ds^2=-dt^2+\sum_{i=1}^{3}C_{i}(t)dx_i^2=-C(\eta)d\eta^2+\sum_{i=1}^{3}C_{i}(t)dx_i^2,\ee
where  $C=(C_1 C_2 C_3)^{1/3}$, $d\eta=dt/C^{1/2}$, and
$u_{\mu}\equiv C^{1/2}(\eta)\delta^{\eta}_{\mu}$. Therefore, in
this frame $F_{\mu\nu}=0$ and $u^{\mu}\nabla_{\mu}u_{\nu}=0$. In
what follows we use primes for denoting derivatives with respect
to the conformal time $\eta$. No sum convention in spatial (latin)
indices is assumed. The generalized dispersion relation takes the
form \be \om^2_k=C(\eta)\lb
m^2+x+2\sum_{s,p}(-1)^{s+p}\,b_{sp}\,x^{(s+p)}\rb, \label{dis} \ee
where
$x=\sum_{i=1}^{3}k_i^2/C_i\equiv\sum_{i=1}^{3}x_i\equiv\sum_{i=1}^{3}
x \lambda_i^2$, with $\sum_{i=1}^3\lambda_i^2=1$.

\section{Self-interacting scalar field in Bianchi type I space-times}
In this section we are concerned with the renormalization of the
equation of motion for the expectation value of a self-interacting
scalar field ($\lambda\neq 0$) propagating in a four-dimensional
Bianchi type I space-time. We assume that the state of the system
is such that the expectation value of the field is $\phi_0$. Then,
defining a new quantum field  $\hat{\phi}$ as
$\phi=\phi_0+\hat{\phi}$, the equation of motion for $\phi_0$ in
the one-loop approximation is given by \be\label{Ecphicero}
\Box\phi_0-\lb m^2+\xi R+\xi_2(\nabla_{\mu}u^{\mu})^2+\xi_3
\nabla_{\mu}u_{\nu}\nabla^{\nu}u^{\mu} +\xi_5
R_{\mu\nu}u^{\mu}u^{\nu}+2\sum_{s,p\leq
s}b_{sp}\mathcal{D}^{2(s+p)}+12\lambda\langle\hat{\phi}^2\rangle\rb\phi_0-4\lambda\phi_0^3=0.\ee
The Fourier modes of the scaled field $\chi=C^{1/2}\hat{\phi}$
satisfy \be {\chi_k''}+\lb \om_k^2+\lp\xi-\frac{1}{6}\rp
CR+Q+\xi_2 C(\nabla_{\mu}u^{\mu})^2+\xi_3 C
\nabla_{\mu}u_{\nu}\nabla^{\nu}u^{\mu}+\xi_5 C
R_{\mu\nu}u^{\mu}u^{\nu}+12 C\lambda
\phi_0^2\rb\chi_k=0,\label{ecparachi}\ee
 with the usual normalization condition \be \ck
{\ck'}^*-\ck'\ck^*=i\; .\label{nor} \ee

The explicit expressions for the different terms in Eqs.
(\ref{Ecphicero}) and (\ref{ecparachi}) are, in Bianchi type I
metrics,
\begin{subequations}
\begin{align}
&(\nabla_{\mu}u^{\mu})^2=\frac{9 D^2}{4 C},\\
&R_{\mu\nu}u^{\mu}u^{\nu}=- \frac{3}{C}\lb\frac{D'}{2}+2Q\rb,\\
&R=\frac{1}{C}\lb 3 D'+\frac{3}{2}D^2+6Q\rb,\\
&\nabla_{\mu}u_{\nu}\nabla^{\nu}u^{\mu}=\sum_{i=1}^3\frac{d_i^2}{4C}=\frac{3}{4C}(D^2+8Q),\\
&Q=\frac{1}{72}\sum_{i<j}^3 (d_i-d_j)^2,\label{Q}
\end{align}
\end{subequations}
where $d_i=C_i'/C_i$ and $D=\sum_{i=1}^3 d_i/3=C'/C$. Note that
for the metric we are considering\be 2
R_{\mu\nu}u^{\mu}u^{\nu}+R=(\nabla_{\mu}u^{\mu})^2-\nabla_{\mu}u_{\nu}\nabla^{\nu}u^{\mu},\ee
and therefore without loss of generality we can set $\xi_5=0$.

For dispersion relations such that the mean value
$\langle\hat{\phi}^2\rangle$ in Eq. (\ref{Ecphicero}) is
divergent, the infinities must be absorbed into the bare constants
of the theory. To implement the renormalization, we start by
expressing the field modes $\chi_k$ in the well known form \be
\ck= \frac{1}{\sqrt{ 2 W_k}}\exp\lp -i\int^\eta
W_k(\tilde\eta)d\tilde\eta\rp, \label{chi} \ee which allows us to
write \be \langle\hat{\phi}^2\rangle =\frac{1}{(2\pi)^3 C}\int
d^3k {|\chi_k|^2}=\frac{1}{(2\pi)^3 C}\int d^3k \frac{1}{2 W_k}.
\ee

Substitution of Eq.
(\ref{chi}) into Eq. (\ref{ecparachi}) yields
\be W_k^2 =
 \om_k^2+\lp\xi-\frac{1}{6}\rp
CR+Q+\xi_2C (\nabla_{\mu}u^{\mu})^2+\xi_3
C\nabla_{\mu}u_{\nu}\nabla^{\nu}u^{\mu}+12\lambda C
\phi^2_0+\frac{5}{16}\frac{[(W_k^2)']^2}{W_k^4}-\frac{1}{4}\frac{(W^2_k)''}{W_k^2}.\label{Weq}
\ee For adiabatic regularization we need the approximate solution
of this non-linear differential equation that is obtained by
assuming that $W^2_k$ is a slowly varying function of $\eta$. In
this adiabatic or WKB approximation the adiabatic order of a term
is given by the number of time derivatives of the metric plus the
power of $\phi_0$ \cite{PazMazzi}. The WKB approximation can be
obtained by  solving the Eq.(\ref{Weq}) iteratively \be W_k =
^{(0)}W_k + ^{(2)}W_k+...\; ,\ee where the superscript denote the
adiabatic order. To lowest order we have $^{(0)}W_k=\om_k$. The
second adiabatic order can be computed replacing $W_k$ by $\om_k$
on the right-hand side of Eq. (\ref{Weq}). Thus,
 we straightforwardly obtain
\begin{eqnarray}
^{(2)}W_k^2 & = & C R(\xi-\frac{1 }{6})+Q+\frac{D^2}{16}-\frac{D'}{4}+\xi_2C (\nabla_{\mu}u^{\mu})^2+\xi_3 C\nabla_{\mu}u_{\nu}\nabla^{\nu}u^{\mu} +12\lambda C \phi^2_0 \nonumber\\
 &-&\frac{(f+1)}{4}\sum_{i=1}^3\lambda_i^2\lb \frac{D d_i}{2}+ d_i^2-d_i' \rb+ \frac{1}{16} \lp\sum_{i=1}^3d_i\lambda_i^2\rp^2 \lb f^2+6 f-4\dot{f}+5\rb,\label{W2}
\end{eqnarray} where we have defined the function
\be f\equiv \frac{d\ln\tilde{\omega}_k^2}{d\ln x}-1,\label{f} \ee
with $\tilde{\omega}_k^2\equiv\omega_k^2/C$. We have also used that
\begin{subequations}\label{derivOmega}
\begin{align}
\frac{(\om_k^2)'}{\om_k^2}=& D-(f+1)\sum_{i=1}^{3}d_i\lambda_i^2, \\
\frac{(\om_k^2)''}{\om_k^2}=&
D'+D^2+(f+1)\sum_{i=1}^{3}\lambda_i^2[d_i^2-2 d_i
D-d_i']+(\dot{f}+f^2+f)\lp\sum_{i=1}^{3}d_i\lambda_i^2\rp^2,
\end{align}\end{subequations} where a dot indicates a derivative with respect to $\ln x$.

We proceed as for the standard dispersion relation, defining the
renormalized expectation value as \be
\langle\hat{\phi}^2\rangle_{ren}=\langle\hat{\phi}^2\rangle-\langle\hat{\phi}^2\rangle_{ad2},
\ee with
$\langle\hat{\phi}^2\rangle_{ad2}=\langle\hat{\phi}^2\rangle^{(0)}+\langle\hat{\phi}^2\rangle^{(2)}$,
where again the superscripts indicate the adiabatic order.

We now compute the zeroth adiabatic order of $\langle \hat{\phi}^2
\rangle$ and regularize it by using the fact that the integral of
a total derivative vanishes in dimensional regularization
\cite{Collins}. For this, and in order to avoid the complications
of computing all quantities in $n$-dimensions, we first perform
the angular integrations and then generalize the four-dimensional
integrals to $n$-dimensions  by replacing
$d^3k=C^{3/2}d^3y=C^{3/2}y^2dy d\Omega$ ($y_i=k_i/\sqrt{C_i}$) by
$C^{3/2}y^{(n-2)}dy d\Omega$.

Therefore, the zeroth adiabatic order is given by \be
\label{rencerophi} \langle\hat{\phi}^2\rangle^{(0)}
=\frac{1}{(2\pi)^3}\int y^{n-2} dy d\Omega \frac{
1}{2\tilde{\om}_k}=\frac{I_1}{2(2\pi)^2}, \ee where $I_1$ is given
Table \ref{tabla}. Note that the integral $I_1$ is divergent
unless $\om_k^2$ behaves as $x^{s}$ with $s>3$, for large values of
$x$. This divergence can be absorbed in the bare mass of the
quantum field (see below).

\begin{table}[ht]
\begin{tabular}{|c|c|}
\hline
& \\
$I_0 =\int_0^\infty dx\,x^{\frac{(n-3)}{2}}{\tilde{\om}_k}$ &
$I_3=\int_{0}^{\infty} dx \frac{x^{\frac{(n-3)}{2}}}{\tilde{\omega}_k^3}$   \\
& \\
\hline
& \\
$I_1=\int_0^\infty dx\,\frac{x^{\frac{(n-3)}{2}}}{\tilde{\om}_k}$&  $I_4=\int_{0}^{\infty} dx \frac{x^{\frac{(n+1)}{2}}}{\tilde{\omega}_k^5} \frac{d^2\tilde{\omega}_k^2}{{dx}^2}$  \\
\hline
& \\
$I_2 = \int_0^\infty
dx\,\frac{x^{\frac{(n+1)}{2}}}{\tilde{\om}_k^3}\frac{d^2\tilde{\om}_k^2}{{dx}^2}$
&
$I=\int_0^{+\infty}dx\frac{x^{\frac{(n+3)}{2}}}{\tilde{\om}_k^3}\frac{d^3\tilde{\om}_k^2}{{dx}^3}$
\\
& \\
\hline
\end{tabular}
 \caption{Explicit expressions for $I_{i}$. To obtain these integrals we have made the change of variables
 $x= y^2$ and we have defined $\tilde{\om}_k=\om_k/\sqrt{C}$.}\label{tabla}
\end{table}

The second adiabatic order can be written as \be
\langle\hat{\phi}^2\rangle^{(2)} =-\frac{\sqrt{C}}{32\pi^3}\int
y^{n-2} dy d\Omega \frac{{}^{(2)}W_k^2 }{\om_k^3}. \ee The angular
integrations can be performed with the use of the identities
listed in the Appendix A. After some calculations we obtain:
\begin{eqnarray}
\langle\hat{\phi}^2\rangle^{(2)}
&=&-\frac{1}{16\pi^2}\left\{I_3\lb\frac{D^2}{16 C}-\frac{D'}{4
C}+R\lp\xi-\frac{1}{6}\rp+\frac{Q}{C}+  \xi_2
(\nabla_{\mu}u^{\mu})^2+\xi_3
\nabla_{\mu}u_{\nu}\nabla^{\nu}u^{\mu} +12\lambda
\phi^2_0\rb\right.\nonumber\\  \label{adPhi} &-&\left.\lb\frac{3
D^2}{8 C}+2\frac{Q}{C}-\frac{D'}{4 C}\rb
(J_{1000}+I_3)+\lb\frac{D^2}{16 C}+\frac{ Q}{5 C}\rb(J_{2000}+6
J_{1000}-4 J_{0100}+5 I_3)\right\},
\end{eqnarray} where $I_3$ is given in Table \ref{tabla}, and we have defined the integrals
\be J _{m n l s}\label{JMNLS}\equiv
\int_{0}^{\infty}dx\frac{x^{\frac{(n-3)}{2}}}{\tilde{\omega}_k^3}
{f}^m\, {\dot{f}}^n\,{\ddot{f}}^l\, {\dddot{f}}^s,\ee with
$m,n,l,s$ integer numbers.  As it is shown in the Appendix of
Ref.\cite{NosDos}, this integrals can be expressed in terms of the
ones in Table \ref{tabla} by performing integrations by parts. For
$n\to 4$, we have \cite{NosDos}:
\begin{subequations}\label{jotas}
\begin{align}
J_{1000}=&0, \\
J_{2000}=&\frac{2}{5}I_4\\
J_{0100}=&\frac{3}{5}I_4.
\end{align}\end{subequations}
Then, substituting this results into Eq. (\ref{adPhi}) we arrive
at
\begin{eqnarray} \label{rendosphi}
\langle\hat{\phi}^2\rangle^{(2)}&=&-\frac{I_3}{16\pi^2}\lb
R\lp\xi-\frac{1}{6}\rp +12\lambda \phi^2_0 +  \xi_2
(\nabla_{\mu}u^{\mu})^2+\xi_3
\nabla_{\mu}u_{\nu}\nabla^{\nu}u^{\mu}
\rb\nonumber\\&+&\frac{I_4}{480\pi^2}\lb (\nabla_{\mu}u^{\mu})^2+2
\nabla_{\mu}u_{\nu}\nabla^{\nu}u^{\mu} \rb.
\end{eqnarray}
This is the main result of this section. The relevant point is
that, in addition of the usual terms proportional to $R$ and
$\phi_0^2$, the second adiabatic order contains terms with two
derivatives of the aether field, which are present even if
$\xi_2=\xi_3=0$. For the standard dispersion relation $I_3$
diverges and $I_4$ vanishes (see Table \ref{tabla}). Therefore,
when $\xi_2=\xi_3=0$ one reobtains the usual result. However,  for
any other dispersion relation of the type given in Eq.
(\ref{dis}), $I_3$ and $I_4$ are finite. An interesting point is
that, if we consider a generalized dispersion relation, evaluate
the integral explicitly in four dimensions and {\it then} take the
limit in which the dispersion relation tends to the usual one,  a
nonvanishing finite result can be obtained. For example, a
dispersion relation of the form $\om_k^2=C(x+2 b_{11}x^2)$ yields
\be I_4=2 b_{11}\int_{0}^{+\infty}dx (1+2
b_{11}x)^{-\frac{5}{2}}=\frac{4}{3}. \ee Therefore,  there is a
finite remnant of the trans-Planckian physics in the second
adiabatic order, even in the limit in which the scale of new
physics is very high $M_C\to\infty$ ($b_{11}\to 0$).

Coming back to the mean value equation (\ref{Ecphicero}), we write
the bare
parameters in terms of the renormalized ones plus the
corresponding to counterterms:
\begin{eqnarray}\label{EcphiceroRe}
&&\Box\phi_0-\left[ m^2_R +\delta m^2+(\xi_R+\delta\xi) R+(\xi_{2R}+\delta\xi_2)(\nabla_{\mu}u^{\mu})^2+(\xi_{3R}+\delta\xi_3) \nabla_{\mu}u_{\nu}\nabla^{\nu}u^{\mu} \right.\nonumber\\
&&+2\sum_{s,p}b_{sp}\mathcal{D}^{2(s+p)}+\left.12\lambda_R\lp\langle\hat{\phi}^2\rangle_{ren}+\langle\hat{\phi}^2\rangle_{ad2}
\rp\right]\phi_0-4(\lambda_R+\delta\lambda)\phi_0^3=0.
\end{eqnarray}
Introducing Eqs. (\ref{rencerophi}) and (\ref{rendosphi})  into
Eq. (\ref{EcphiceroRe}), we see that the regularized second
adiabatic order $\langle\hat{\phi}^2\rangle_{ad2}$ can be absorbed
into the bare constants by defining counterterms such that
\begin{subequations}\label{contraterminos}
\begin{align}
\delta m^2=&-\frac{6\lambda_R }{(2\pi)^2}I_1, \\
\delta\lambda=&\frac{9\lambda_R^2}{(2\pi)^2} I_3, \\
\delta\xi=&\frac{3\lambda_R}{(2\pi)^2}\lp\xi_R-\frac{1}{6}\rp I_3,\\
\delta\xi_2=&\frac{3\lambda_R\xi_{2R}}{(2\pi)^2} I_3-\frac{\lambda_R}{40\pi^2}I_4, \\
\delta\xi_3=&\frac{3\lambda_R\xi_{3R}}{(2\pi)^2}
I_3-\frac{\lambda_R}{20\pi^2}I_4.
\end{align}\end{subequations}
Note that, even  when the parameters $\xi_{2R}$ and $\xi_{3R}$ are set to zero, the corresponding counterterms
arise due to the self-interaction of the scalar field. Note also that by considering the same theory but in a background
flat FRW space-time, it is not possible to distinguish between the
redefinitions of $\xi_2$ and $\xi_3$ proportional
to $I_4$, since in such background we have that \be
\nabla_{\mu}u_{\nu}\nabla^{\nu}u^{\mu}=\frac{1}{3}(\nabla_{\mu}u^{\mu})^2.\ee

From the results of this section we conclude that, as long as one
considers the renormalization of the mean value equation in
Bianchi type I spacetimes, and for the class of MDR considered
here, it is enough to subtract the zeroth adiabatic order of
$\langle\hat\phi^2\rangle$, since the second adiabatic order
produce a finite renormalization of the bare constants of the
theory. It would be interesting to check whether this is a general
property, i.e. valid for an arbitrary background, or not. In order
to address this issue, it would be necessary to know the
singularity structure of the two-point function of a quantum field
satisfying MDR for arbitrary values of $g_{\mu\nu}$ and $u_{\mu}$.
This singularity structure could be revealed by a generalized
momentum-space representation of the Green's functions
\cite{bp79,r07}. In any case, the calculation of the second
adiabatic order presented in this section shows that the
interaction terms proportional to $\xi_2$ and $\xi_3$ that appear
in Eq. (\ref{lintgen}) are generated by quantum effects, even if
not present at the classical level. It is likely that the other
interaction terms will also be generated in a more general
background.

\section{On the renormalization of the stress tensor in Bianchi type I space-times}

In this section we  focus on the renormalization of the SEAE. We
restrict the analysis to the case of a free scalar field
($\lambda=0$, $\langle\phi\rangle=0$) and, for the sake of
simplicity, we set the parameters $\xi_i=0$ (i=1,2,3,4,5).

The SEAE take the form \be G_{\mu\nu}+\Lambda g_{\mu\nu}=8\pi G
[T_{\mu\nu}^{u_b}+\langle
T_{\mu\nu}^{\tilde{\lambda}_{c}}+T_{\mu\nu}^{\phi}\rangle +
T_{\mu\nu}^{clas}] \ee where $\Lambda$ and $G$ are the bare
cosmological and Newton's constants, $G_{\mu\nu}$ is the Einstein
tensor, $T_{\mu\nu}^{u,(\phi)}=-\frac{2}{\sqrt{-g}}\frac{\delta
S^{u(\phi)}}{\delta g^{\mu\nu}}$, and $T_{\mu\nu}^{u}=
T_{\mu\nu}^{u_b}+T_{\mu\nu}^{\tilde{\lambda}_{c}}$,  $
T_{\mu\nu}^{u_b}$ is the stress tensor of the background vector
field while $T_{\mu\nu}^{\tilde{\lambda}_{c}}$ is the additional
contribution due to the modification of the Lagrange multiplier
$\tilde{\lambda}$ arising from the coupling between the scalar
field $\phi$ and  $u_{\mu}$. $T_{\mu\nu}^{clas}$ is a stress
tensor coming from classical sources not coupled to the aether
field. As we will compute the mean value of the stress tensor  up
to the second adiabatic order, we omit classical terms quadratic
in the curvature (we will comment on this issue in the next
section).

The nontrivial components of the Einstein tensor are, in Bianchi
type I spacetimes:
\begin{subequations}
\begin{align}
G_{\eta\eta}=& 3\lb\frac{D^2}{4}-Q\rb, \\
G_{ii}=&- \frac{C_i}{2C}\lb 3 D'+\frac{3}{2}D^2+6 Q-d_i'-d_i D\rb.
\end{align}\label{EinsteinTensorBianchi}
\end{subequations}
The stress tensor corresponding to the background vector field
$u_{\mu}$ can be written as \be T_{\mu\nu}^{u_b}=-\frac{b_3}{8\pi
G}G_{\mu\nu}-\frac{b_2}{8\pi G}\tilde{T}_{\mu\nu}^{u}, \ee whose
nonzero components  are
\begin{subequations}
\begin{align}
\tilde{T}_{\eta\eta}^{u}=& \frac{9 }{8} D^2, \\
\tilde{T}_{ii}^{u}=& -\frac{3}{2}\frac{C_i}{C}\lb
D'+\frac{D^2}{4}\rb.
\end{align}\label{AETHERTensorBianchi}
\end{subequations}

The expectation value of the quantum energy momentum tensor
$T_{\mu\nu}=T_{\mu\nu}^{\phi}+T_{\mu\nu}^{\tilde{\lambda}_c}$ is
given by
\begin{eqnarray}
\nonumber\langle T_{\eta\eta}\rangle &=& \frac{1}{2 C} \int
\frac{d^{3}k}{(2\pi)^{3}} \left\{
|\chi_k'|^2+3 D \lp\xi-\frac{1}{6}\rp (\chi_k'\chi_k^*+\chi_k{\chi_k'}^{*})\right.\\
&+&\left.|\chi_k|^2\lb\om_k^2-3D^2\lp\xi-\frac{1}{12}\rp+2\xi G_{\eta\eta}\rb \right\},\\
\nonumber \langle T_{ii}\rangle  &=&  \frac{C_i}{C^2}\int
\frac{d^3 k}{(2\pi)^{3}}
\left\{\lp\frac{1}{2}-2\xi\rp |\chi_k'|^2+\lp\frac{\xi}{2}(2 D+d_i)-\frac{D}{4}\rp (\chi_k'\chi_k^*+\chi_k{\chi_k'}^{*})\right.\\
 &-&\left.  \xi(\chi_k''\chi_k^*+\chi_k{\chi_k''}^{*})+|\chi_k|^2\lp k_i^2\frac{d\om_k^2}{
dk_i^2}-\frac{\om_k^2}{2}+\frac{D^2}{8}\rp+\frac{\xi}{2}
|\chi_k|^2\lb 2 D'-d_i D+2\frac{C}{C_i} G_{ii}\rb\right\}.
\end{eqnarray}
Using the expression given in Eq. (\ref{chi}) for the modes
$\chi_k$, it can be written as
\begin{eqnarray}
\nonumber\langle T_{\eta\eta}\rangle &=& \frac{1}{2 C} \int
\frac{d^{3}k}{(2\pi)^{3}} \left\{
\frac{[(W_k^2)']^2}{32 W_k^5}-3 D\lp\xi-\frac{1}{6}\rp\frac{(W_k^2)'}{4 W_k^3 }+\frac{W_k}{2}\right.\\
&+&\left.\frac{1}{2 W_k}\lb\om_k^2-3D^2\lp\xi-\frac{1}{12}\rp+2\xi G_{\eta\eta}\rb\right\},\label{Tetaeta}\\
\nonumber \langle T_{ii}\rangle  &=&  \frac{C_i}{C^2}\int
\frac{d^3 k}{(2\pi)^{3}}
\left\{\lp\frac{1}{8}-3\xi\rp \frac{[(W_k^2)']^2}{8 W_k^5}+\xi \frac{(W_k^2)''}{4 W_k^3}-\frac{(W_k^2)'}{4 W_k^3}\lp-\frac{D}{4}+\frac{\xi}{2}(2 D+d_i)\rp\right.\\
 &+&\left. \frac{W_k}{4}+\frac{1}{2 W_k}\lb k_i^2\frac{d\om_k^2}{
dk_i^2}-\frac{\om_k^2}{2}+\frac{D^2}{8}+\xi D'-\frac{\xi}{2}D
d_i+\xi\frac{C}{C_i} G_{ii}\rb\right\}.\label{PPP}
\end{eqnarray}

Therefore, the zeroth adiabatic order can be expressed in the form
\begin{subequations}
\begin{align}
\langle T_{\eta\eta}\rangle^{(0)} =& \frac{C}{2} \int
\frac{d\Omega dy}{(2\pi)^{3}}
y^{n-2}\tilde{\om}_k=\frac{C}{2(2\pi)^2}\int_0^{+\infty}dx
x^{\frac{n-3}{2}}\tilde{\om}_k,\\\label{ordenceroii} \langle
T_{ii}\rangle^{(0)}=&  \frac{C_i}{2 } \int \frac{d\Omega
dy}{(2\pi)^{3}} y^{n-2} \lambda_i^2
\frac{y^2}{\tilde{\om}_k}\frac{d\tilde{\om}_k^2}{dy^2}=\frac{
C_i}{3 (2\pi)^2}\int_{0}^{+\infty}dx
x^{\frac{n-1}{2}}\frac{d\tilde{\om}_k}{dx},
\end{align}
\end{subequations}where we have used that $\int d\Omega \lambda_i^2=4\pi/3$.
Then, after an integration by parts in Eq. (\ref{ordenceroii}) we
obtain, as $n\to4$, \be \langle
T_{\mu\nu}\rangle^{(0)}=-\frac{I_0}{2(2\pi)^2}g_{\mu\nu},\ee where
$I_0$ is a divergent integral as $n\to 4$ for any of the
dispersion relations given in Eq. (\ref{dis}) (see Table
\ref{tabla}). Hence, this regularized adiabatic order can be
absorbed into a redefinition of the bare cosmological constant
$\Lambda$.

The second adiabatic order of $\langle T_{\mu\nu}\rangle$ can be
written  as
\begin{eqnarray}
\langle T_{\eta\eta}\rangle^{(2)} &=& \frac{C}{2} \int
\frac{d\Omega dy}{(2\pi)^{3}}
\frac{y^{(n-2)}}{\tilde{\om}_k}\left\{
\frac{[(\om_k^2)']^2}{32 \om_k^4}-3 D\lp\xi-\frac{1}{6}\rp\frac{(\om_k^2)'}{4 \om_k^2}-\frac{3}{2}D^2\lp\xi-\frac{1}{12}\rp+\xi G_{\eta\eta}\right\},\label{TetaetasecOeden}\\
\nonumber \langle T_{ii}\rangle^{(2)}  &=&  C_i\int \frac{d\Omega
dy}{(2\pi)^{3}}\frac{y^{(n-2)}}{\tilde{\om}_k}
\left\{\lp\frac{1}{8}-3\xi\rp \frac{[(\om_k^2)']^2}{8 \om_k^4}+\xi \frac{(\om_k^2)''}{4 \om_k^2}-\frac{(\om_k^2)'}{4 \om_k^2}\lp-\frac{D}{4}+\frac{\xi}{2}(2 D+d_i)\rp\right.\\
 &+&\left.\frac{^{(2)}W_k^2}{4}\lp1-\lambda_i^2\frac{y^2}{\om_k^2}\frac{d\om_k^2}{dy^2}\rp+\frac{D^2}{16}+\frac{\xi}{2} D'-\frac{\xi}{4}D d_i+\xi\frac{C}{2 C_i} G_{ii}\right\},\label{PPPSegorden}
\end{eqnarray} where ${}^{(2)}W_k^2$ is given by the expression in Eq. (\ref{W2}) with $\lambda=\xi_{2}=\xi_{3}=0$. The explicit expressions for  $(\om_k^2)'/{\om}_k^2$ and $({\om}_k^2)''/{\om}_k^2$ are given in Eq. (\ref{derivOmega}).

After performing the angular integrations with the use of the
identities given in the Appendix A and some algebraic
manipulations, we obtain:

\begin{subequations}\label{segordenInt}
\begin{align}
\langle T_{\eta\eta}\rangle^{(2)} &= \frac{1}{(2\pi)^2}[\alpha_1 D^2+\alpha_2 Q],\\
 \langle T_{ii}\rangle^{(2)}  &=  \frac{C_i}{C(2\pi)^2}[\beta_1 D^2+\beta_2 D'+\beta_3 D d_i+\beta_4 Q+\beta_5 d_i^2+\beta_6
 d_i'].
\end{align}
\end{subequations}
The coefficients $\alpha_i$ and $\beta_i$ are given in Appendix B,
where  it is also shown that using integration by parts they can be expressed in terms of two
of the integrals in Table \ref{tabla}. Thus, we find
\be\label{divad2tmunu} \langle
T_{\mu\nu}\rangle^{(2)}=\frac{1}{8\pi^2}\left\{\lb I_1\lp
\xi-\frac{1}{6}\rp -\frac{I_2}{45}\rb G_{\mu\nu}+\frac{I_2}{30}
\tilde{T}^{u}_{\mu\nu}\right\}.\ee

Note that both $I_1$ and $I_2$ diverge when $\om_k^2$ behaves as
$x^{s}, s\leq 3$ for large values of $x$. Therefore, in this case
the divergences should be absorbed into the bare constants $G$ and
$b_2$. However, when $s>3$, the second adiabatic order produce
finite renormalizations of both constants.

As in the evaluation of $\langle\hat\phi^2\rangle$ presented in
the previous section, depending on the dispersion relation one
could have a remnant of the trans-Planckian physics in the second
adiabatic order of $\langle T_{\mu\nu}\rangle$. Indeed, while
$I_2$ vanishes for the standard dispersion relation, a non
vanishing (and even divergent) result can be obtained for MDR in
the limit $M_C\to\infty$. For example, for a dispersion relation
of the form $\om_k^2=C(x+2 b_{22} x^4)$ we find that \be I_2=24
b_{22}\int_{0}^{+\infty}dx \frac{x^3}{(1+2 b_{22}
x^3)^{\frac{3}{2}}}=\frac{2^{\frac{8}{3}}}{\sqrt{\pi}b_{22}^{\frac{1}{3}}}\Gamma\lb1/6\rb\Gamma\lb4/3\rb,
\ee which diverges as $M_C\to\infty$ ($b_{22}\to 0$).

Eq. (\ref{divad2tmunu}) is the main result of this section. We see
that, for a generalized dispersion relation of the type given in
Eq. (\ref{dis}), not only a redefinition of the Newton's constant
is necessary  in order to cancel the divergences of the second
adiabatic order,  but also a redefinition of the coefficient $b_2$
which corresponds to the term $(\nabla_{\mu}u^{\mu})^2$ in the
bare Lagrangian of the vector field. The second adiabatic order
contains terms that are non-purely geometric, in the sense that
they cannot be written only in terms of the metric, but also
involve the aether field.

It is noteworthy that for a background flat FRW space-time
$G_{\mu\nu}=3/2 \tilde{T}^{u}_{\mu\nu}$ , thereby, in
Refs.\cite{NosUno,NosDos} it was not possible to realize that a
redefinition of the Newton's constant is not enough for cancelling
the second adiabatic order. In fact, for this particular
space-time $G_{\mu\nu}$ is the unique covariantly conserved tensor
of adiabatic order two that can be derived from an action formed
by combining the vector field $u^{\mu}$, the metric $g_{\mu\nu}$,
and their derivatives.

\section{Discussion}

In this paper we have worked within the context of  a generally
covariant theory of gravitation coupled to a dynamical time-like
Lorentz-violating vector field. We considered a quantum scalar
field satisfying MDR, and analyzed the renormalization of the
infinities that arise in the semiclassical theory. In particular,
considering  Bianchi type I spacetimes,  we have analyzed the
dynamical equation for the expectation value of a self-interacting
scalar field (Section III), and the SEAE for the metric in the
case of a free scalar field (Section IV). With the use of
adiabatic subtraction and dimensional regularization, we have
shown that, in addition to the usual terms required to absorb the
infinities of the second adiabatic orders, it is necessary to
consider more general counterterms that involve the aether field.
This property was not apparent in our previous works
\cite{NosUno,NosDos}, due to the high symmetry of the flat
Robertson Walker metrics.

These results suggest that, in a more general background metric,
any covariant term  which can be formed by combining the vector
field $u^{\mu}$, the metric $g_{\mu\nu}$ and up to two of their
derivatives, will appear in the regularized second adiabatic order
of the expectation value of the quantum stress tensor, provided
that the theory contains a scalar field with a generalized
dispersion relation of the type given in  Eq. (\ref{dis}). Hence,
in order to absorb the divergences contained in the second
adiabatic order, a bare action as general as the one given in Eq.
(\ref{Sg}) should be considered. Depending on the particular
dispersion relation of the quantum field, the second adiabatic
order may be finite. If this is the case, quantum effects generate
finite renormalizations of the constants appearing in the
classical Lagrangian. As we have also pointed out in Section IV,
this finite renormalizations could be extremely large.

In the weak-field limit, the terms proportional to the constants
$b_i$ in Eq. (\ref{lu})  could have observable consequences. Indeed,
the most general action given in Eq. (\ref{Sg}) has four free
parameters more than general relativity. This theory has been
studied in several contexts, such as of the static weak-field
limit \cite{Jacobsondebil}, the radiation and propagation of the
aether-gravitational waves \cite{Jacobsonwave},  cosmology
\cite{LimCarrolBarrow}, etc., in which stringent constraints on
the parameters have been imposed to make the theory consistent
with observation. For example, in Ref. \cite{Jacobsondebil} it is
shown that for all the PPN parameters to agree with observation,
the four additional  parameters of the model must satisfy two
constraint equations with sufficient accuracy (i.e., the
additional four-parameter space of the model has to be practically
reduced to a two-dimensional subspace).  In the absence of a known
mechanism to explain why the parameters satisfy precisely such
constraint equations, it seems that quantum effects generate a
fine-tuning problem in the Einstein-Aether theory. This is analogous to
the fine-tuning problem present in the Myers-Pospelov modification
of QED \cite{pipi}.

In this paper we restricted ourselves to the evaluation of the
adiabatic expansion up to the second adiabatic order, in a
particular class of background metrics. By power counting, we
expect the fourth adiabatic order $\langle
T_{\mu\nu}\rangle^{(4)}$ to be finite in these metrics. However,
there could be some subtleties related to the would be
Gauss-Bonnet invariant in four dimensions \cite{NosProc,NosDos}.
In the light of the results obtained in this paper, this issue
should be reexamined. One should compute $\langle
T_{\mu\nu}\rangle^{(4)}$ for a general background metric and
aether field (this could be done by generalizing the
momentum-space representation of the Green's functions). On
dimensional grounds we expect $\langle T_{\mu\nu}\rangle^{(4)}$ to
contain terms proportional to the variation of $R
(\nabla_{\mu}u^{\mu})^2$, $(R_{\mu\nu}u^{\mu}u^{\nu})^2$,
$R_{\mu\nu}u^{\mu}u^{\nu} R$, $R_{\mu\nu\rho\sigma}u^{\mu}u^{\rho}
R^{\nu\sigma}$, etc., in addition to the usual ones: $R^2$,
$R_{\mu\nu}R^{\mu\nu}$ and
$R_{\mu\nu\rho\sigma}R^{\mu\nu\rho\sigma}$. Depending on the MDR,
the fourth adiabatic order could be finite or divergent when
expressed in terms of such variations. This fact would define
whether it is necessary or not to subtract the fourth adiabatic
order in a general background, for a given dispersion relation.
Work in this direction is in progress.

\section*{Appendix A: Identities for  Bianchi type I space-times}

In this Appendix we briefly summarize some useful formulas
required for the adiabatic regularization of $\langle
\hat{\phi}^2\rangle^{(2)}$ and $\langle T_{\mu\nu}\rangle^{(2)}$
in  Bianchi type I space-times.

As we have already mentioned in the text, in order to regularize
the theory we perform the four-dimensional angular integrations
and then generalize the integrals to $n$-dimensions. We rescale
the integration variables $k_i\to y_i=k_i/C_i$ and transform the
volume element $d^{3}y$ from rectangular coordinates to spherical
coordinates $y^2 dy d\Omega$, where $d\Omega$ is the solid angle
element. In terms of $y^2_i=y^2 \lambda_i^2$, the relevant
integrals are of the form \be I(i,j,k)= \int d\Omega
\lambda_1^{2i}\lambda_2^{2j}\lambda_3^{2k}, \ee which can be
evaluated by using the fact that they are invariant under
permutations of $\{i,j,k\}$. We provide here a list of the
integrals we have used in this paper (see \cite{Hu} for more
details):
\begin{subequations}\label{ANGULARINT}
\begin{align}
I(0,0,k)=&\frac{4\pi}{2k+1},\\
I(1,1,0)=&\frac{4\pi}{5\times 3},\\
I(1,2,0)=&\frac{4\pi}{7\times 5},\\
I(1,1,1)=&\frac{4\pi}{7\times 5 \times 3}.
\end{align}
\end{subequations}
These results, together with the formula $\sum_{i=1}^3d_i^2=3(8
Q+D^2)$, allow us to  derive the following  identities:
\begin{subequations}\label{identities}
\begin{align}
\sum_{j=1}^{3}\sum_{k=1}^{3}\int d\Omega d_j d_k \lambda_j^2\lambda_k^2=&4\pi\lp D^2+\frac{16}{5}Q\rp,\\
\sum_{j=1}^{3}\sum_{k=1}^{3}\int d\Omega d_j d_k \lambda_i^2\lambda_j^2\lambda_k^2=&\frac{4\pi}{7\times 5}\lb 5 D^2+4 d_i D +\frac{8}{3}d_i^2+16 Q\rb,\\
\sum_{j=1}^{3}\int d\Omega (d_j'+2 d_j D-d_j^2)
\lambda_i^2\lambda_j^2=&\frac{4\pi}{5\times 3}\lb 2(d_i'+2 d_i
D-d_i^2)+3(D'+D^2-8 Q)\rb\; ,
\end{align}
\end{subequations}
that are useful for the evaluation of $\langle
\hat{\phi}^2\rangle^{(2)}$ and $\langle T_{\mu\nu}\rangle^{(2)}$.

\section*{Appendix B: Regularization of $\langle T_{\mu\nu}\rangle^{(2)}$ in  Bianchi type I space-times}
In this Appendix we provide some details for computing the second
adiabatic order of the expectation value of the quantum energy
momentum tensor. The explicit expressions for the coefficients
appearing in Eq. (\ref{segordenInt})are:
\begin{subequations}\label{coeff}
\begin{align}
\alpha_1=&\frac{1}{64}[-4 I_{10}+ I_{20}+4( I_{00}-6\xi I_{00}+6\xi I_{10})],\\
\alpha_2=&\frac{1}{20}[ I_{00}+2 I_{10}+ I_{20}-30\xi I_{00}],\\
\beta_1=&\frac{1}{2240}[52 I_{00}-120 I_{01}-76 I_{10}+20 I_{11}+77 I_{20}-5 I_{30}+280\xi (-I_{00}+2 I_{01}+I_{10}-I_{20})],\\
\beta_2=&\frac{1}{640}[2 I_{00}+39 I_{10}-8 I_{20}-20\xi(22 I_{00}+I_{10})],\\
\beta_3=&\frac{1}{1680}[-8 I_{00}+12 I_{01}-19 I_{10}+12 I_{11}-14 I_{20}-3 I_{30}+210\xi I_{10}],\\
\beta_4=&\frac{1}{140}[-24 I_{01}+3 I_{10}+4 I_{11}+21 I_{20}-I_{30}-14\xi(-8 I_{01}+I_{10}+4 I_{20})+I_{00} (42\xi-19)],\\
\beta_5=&\frac{1}{840}[2 I_{00}+4 I_{01}+3 I_{10}+4 I_{11}-I_{30}],\\
\beta_6=&\frac{1}{120}[-I_{00}-2 I_{10}-I_{20}+30\xi I_{00}],
\end{align}
\end{subequations}
where the integrals $I_{mn}$ are given by \be
I_{mn}=\int_{0}^{+\infty} dx
\frac{x^{\frac{n-3}{2}}}{\tilde{\om}_k} f^{m}\dot{f}^n, \ee  with
$m,n=0,1,2,3$.

Let us now sketch the procedure to find relations between these
coefficients in the context of dimensional regularization, which
is completely analogous to the one described in the Appendix of
Ref.\cite{NosDos} for relating the integrals $ J _{m n l s}$ of
Eq. (\ref{JMNLS}). By definition, $I_{00}=I_1$, and
\begin{eqnarray}
I_{10}&=&\int_{0}^{+\infty} dx \frac{x^{\frac{n-3}{2}}}{\tilde{\om}_k}\lp\frac{x}{\tilde{\om}_k^2}\frac{d\tilde{\om}_k^2}{dx}-1\rp=-2\int_{0}^{+\infty} dx x^{\frac{n-1}{2}}\frac{d\tilde{\om}_k^{-1}}{dx}-I_1\\\nonumber
&=&(n-1)I_1-I_1 \,{ }_{\overrightarrow{(n\to 4)}} 2I_1,
\end{eqnarray}where we have performed an integration by parts and discarded the surface term.
Similarly, one can prove that
\begin{subequations}
\begin{align}
I_{20}=&\frac{2}{3}I_2,\\
I_{30}=&\frac{4}{5}I_2+\frac{8}{15}I,\\
I_{01}=&-2I_1+\frac{1}{3}I_2,\\
I_{11}=&-\frac{2}{15}I_2+\frac{2}{15}I,
\end{align} where the integrals $I_i$ on the right hand side are given in Table \ref{tabla}, and $I$ (which does not appear in the final results) is given by
\be
I=\int_{0}^{\infty} dx \frac{x^{\frac{(n+3)}{2}}}{\tilde{\om}_k^3} \frac{d^3\tilde{\om}_k^2}{dx^3}.
\ee

\end{subequations}
Replacing these results into Eq. (\ref{coeff}), we obtain:
\begin{subequations}
\begin{align}
\alpha_1=&\frac{3}{8}I_1\lp\xi-\frac{1}{6}\rp+\frac{I_2}{96},\\
\alpha_2=&-\frac{3}{2}I_1\lp\xi-\frac{1}{6}\rp+\frac{I_2}{30},\\
\beta_1=&-\frac{3}{8}I_1\lp\xi-\frac{1}{6}\rp+\frac{I_2}{480},\\
\beta_2=&-\frac{3}{4}I_1\lp\xi-\frac{1}{6}\rp-\frac{I_2}{120},\\
\beta_3=&\beta_6=\frac{1}{4}I_1\lp\xi-\frac{1}{6}\rp-\frac{I_2}{180},\\
\beta_4=&-\frac{3}{2}I_1\lp\xi-\frac{1}{6}\rp+\frac{I_2}{30},\\
\beta_5=& 0.
\end{align}
\end{subequations}
Finally, after substituting these coefficients into Eq.
(\ref{segordenInt}) we arrive at Eq. (\ref{divad2tmunu}).
\begin{acknowledgments}
We would like to thank T. Jacobson and H. Vucetich for useful correspondence and
discussions.
This work has been supported by  Universidad de Buenos Aires,
CONICET and ANPCyT.

\end{acknowledgments}

\end{document}